**Shuyi: A Name After Dendritic Cell-mediated Immunological Memory**

**Yiqiang Wang, PhD, Professor**

The First Affiliated Hospital of Soochow University, Soochow University, Suzhou, China.
[Current address: Central Lab, Xiang'an Hospital of Xiamen University, Xiamen University, Xiamen, China].
Correspondence: yiqiangwang99@hotmail.com

Immunological memory is a fundamental theory of modern immunology, which is traditionally believed to be mediated only by B and T lymphocytes that recognize antigen epitopes in a receptor-restricted manner. During the last decade data accumulated to show that monocytes and macrophages, the two main initiators of innate immune response, also built up a "memory" to antigens they encountered, though in most concerned publications a different wording (i.e. "train" or "educate") was utilized to describe this feature [1, 2, 3, 4, 5, 6]. More recently, Hole et al demonstrated a "memory-like" response of dendritic cells (DCs) [7]. In brief, if fungal-challenged mice could develop a protective immune response, DCs immediately (in 3 weeks) isolated from those mice would manifest a pro-inflammatory phenotype. Even after the mice were allowed to rest for 10 weeks, DCs from them still exhibited an enhanced immune activation profile in their transcriptome and cytokine productions upon re-challenge with same pathogens. Lastly, Hole showed that the "training" or memory-building in DCs was achieved by histone modification [7]. All above findings obtained in monocytes, macrophages or DCs emphasized the necessity for rechecking the questions whether antigen presenting cells (APCs) as a whole could be classified the third class of cells that would mediate immunological memory.

During my postdoctoral training and early career in China in late 1990s, DCs-related topics were spotlights of immunology. The lab I worked in was initiating several exciting projects in this field, such as DCs-based vaccines [8], transdifferentiation of leukemic cells into DCs [9], or hypothetical regulatory DCs [10, 11]. Based on the plain logic "the earlier the better" for any protection system like the immune system, I proposed that if DCs, or any other APCs, would build up memory of an antigen they encounter, process, and present, it would be more economical than a memory built-up at later stages such as those mediated by T and B cells. I gained support of my supervisor Xuetao Cao and worked on this hypothesis. As a beginning, we utilized severe combined immunodeficiency (SCID) mice to avoid interference of T or B cells-mediated memory against an antigen. In brief, the Balb/c-SCID mice were challenged with either two doses of soluble antigen ovalbumin (OVA, with aluminum hydroxide as adjuvant) via footpad injection, and then rested for three months, allowing primary immune responses to subside. Then splenic lymphocytes prepared from naïve Balb/c mice were adoptively transferred into the primed or age-control SCID mice (day 0). The mice started to produce anti-OVA IgG in serum around day 12 (i.e. later than regular timeline) even without being challenged with OVA. More surprisingly, the anti-OVA titers in sera manifested a pattern mimicking secondary immune response, namely increasing faster and reaching higher than in paralleling primary immunizing controls. We believed that it was the APCs compartment of SCID mice primed three months earlier in absence of any functional T and B cells gained and maintained the ability to activate antigen-specific T cells in a more efficient way for at least three months. To test if this apparent memory was actually due to potential residual OVA antigens administered three months earlier, OVA was labeled with isotope $^{125}$I and used for priming SCID mice. Three months later, the amount of residual antigens in injection site (whole footpad) as calculated by measuring $^{125}$I radioactivity was minimal and below the amount necessary to initiate a detectable immune response if used for challenging naïve immunocompetent mice directly. Furthermore, when OVA-primed and naïve T/B cells-reconstituted SCID mice were subjected to OVA re-challenge three days after reconstitution, a secondary pattern immune response was observed, again reflected by sera anti-OVA IgG titer curves. On the contrary, when non-OVA-primed but similarly reconstituted SCID mice were similarly challenged by OVA, they developed a primary pattern immune response as expected.

For explaining the memory-like status observed in OVA-primed SCID mice, we further proposed that the APCs were activated by priming antigens and, even in absence of capable T/B lymphocytes and independent of priming antigens, could live for at least three months and, when encountering naïve T/B lymphocytes, were able to present their antigenic epitopes loads to lymphocytes efficiently, i.e. in a pattern more efficiently than newly challenged APCs. To check these possibilities, we isolated nucleated splenocytes, bone marrow cells, or draining lymph node cells from OVA-primed SCID mice three months later, and transferred them respectively into naïve Balb/c mice, which were then challenged with OVA three days later. It was found that splenic and lymph node cells enabled the recipient mice a memory-like response, but bone marrow cells did not. Furthermore, when above splenocytes were depleted of adherent cells (supposedly most APCs) before adoptive transfer, their ability to initiate a memory-type response in naïve recipients was greatly inhibited. Finally, pure DCs were cultured from naïve bone marrow cells in vitro, primed with OVA and then transferred into SCID mice. Again, three months later, when the mice were reconstituted with naïve Balb/c splenocytes and challenged with OVA, they produced higher anti-OVA IgG titers than mice previously transferred with non-primed DCs.

While the idea of DCs mediating immunological memory gradually became clear and gained partial experimental supports, my first baby was conceived and developed in about the same timeline. When she was born in 1997, I named her Shuyi to mark those exciting days and the enthusiasm I put on that project. In Chinese language of this context, Shuyi is for the short form or abbreviation of "dendritic cells-mediated immunological memory". That story was recorded in the report I submitted when I finished my postdoc training and took a lecturer position in that university in 1998 (Figures 1, 2). Regretfully, the manuscript based on above data of APCs-mediated "immunological memory" was rejected by a top journal and had never been tried with any other journals, though results of the other two projects (Figure 1) I worked on during that period were published later [12, 13]. Afterwards, my research shifted into other field and that "DC-memory" project was kind of discontinued. Still, I had been encouraged by recent progresses made worldwide in this field. In my opinion, with the great leap in the field of mice engineering and single cell sequencing or other high-throughput technologies in last decades, it is feasible now to demonstrate the existence of memory-type DCs with more convincingness. For example, large number of DCs could be expanded from bone marrow of mice bearing specific biomarker (e.g. fluorescent GFP), primed with OVA and transferred into syngenic $GFP^{-}$-SCID or immunocompetent mice. If memory type-DCs do exist and survive for three months, it would be practical to sort such $GFP^{+}$-DCs out from hypothetical residing organs, such as spleens, draining lymph nodes, or bone marrow. In case those memory-type DCs would be able to proliferate in response to re-challenging antigens, the chance to identify them in tissues or to sort them out from their residing organs would further increase when the animals are re-challenged at desired time points. Then the powerful single-cell sequencing or any other powerful techniques like the used by Hole study [7] could be utilized to identify the molecular features of these re-activated memory-like DCs. Surely, comparison of such features among re-activated memory-like DCs, resting (but GFP+) OVA-primed DCs and naïve DCs, etc, would reveal clues to the mechanisms for DCs to mediate immunological memory.

## Competing interests statement

The author has no competing interests.

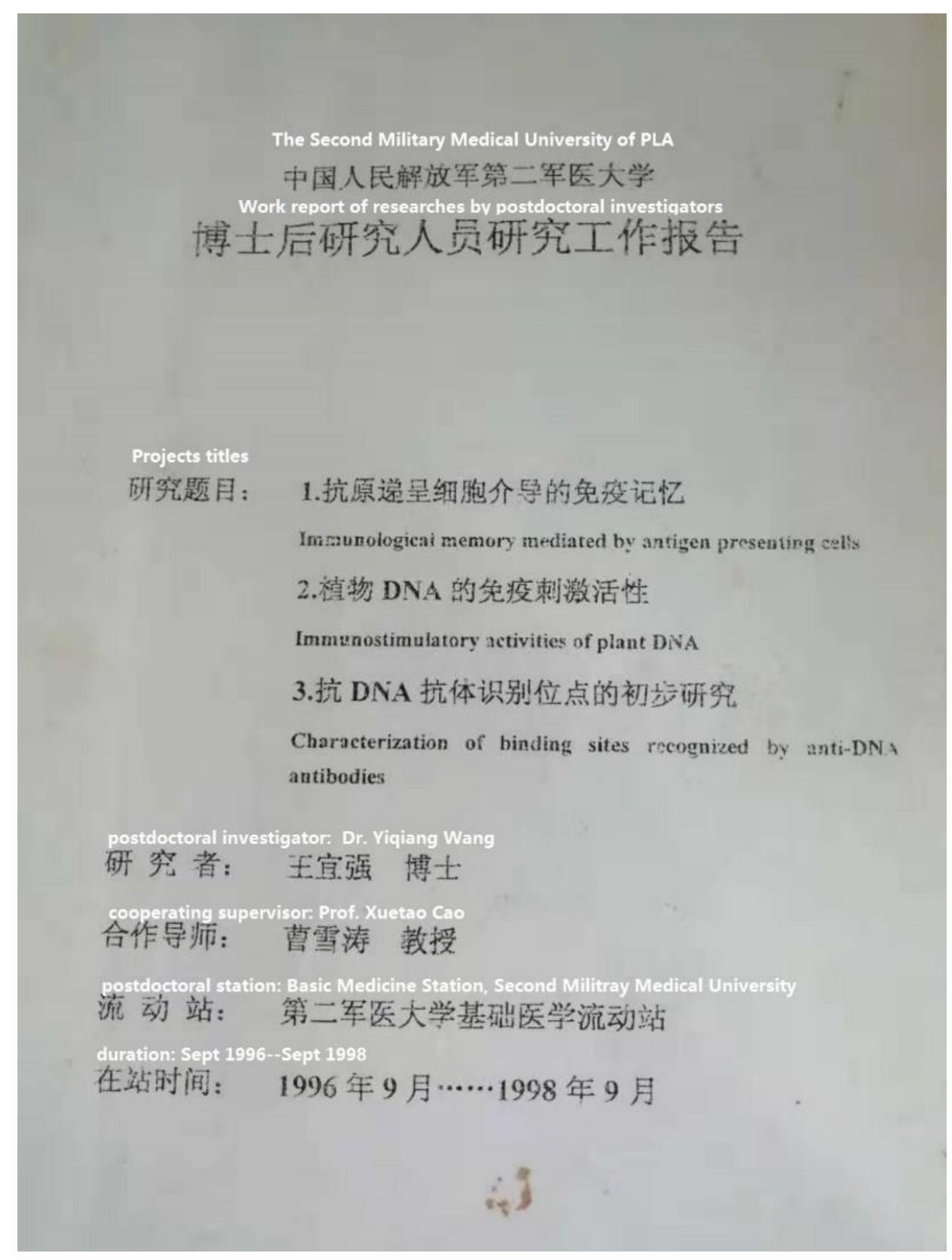

The Second Military Medical University of PLA
中国人民解放军第二军医大学
Work report of researches by postdoctoral investigators
博士后研究人员研究工作报告

Projects titles
研究题目： 1.抗原递呈细胞介导的免疫记忆
Immunological memory mediated by antigen presenting cells
2.植物 DNA 的免疫刺激活性
Immunostimulatory activities of plant DNA
3.抗 DNA 抗体识别位点的初步研究
Characterization of binding sites recognized by anti-DNA antibodies

postdoctoral investigator: Dr. Yiqiang Wang
研 究 者： 王宜强 博士
cooperating supervisor: Prof. Xuetao Cao
合作导师： 曹雪涛 教授
postdoctoral station: Basic Medicine Station, Second Militray Medical University
流 动 站： 第二军医大学基础医学流动站
duration: Sept 1996--Sept 1998
在站时间： 1996 年 9 月……1998 年 9 月

**Figure 1. Picture of the cover of Wang's report for ending his postdoctoral training in September 1998.** After about 21 years since the production of the report book, the black prints on the cover are somewhat blurring but still readable. The titles of the three worked-on projects were originally provided bilingually. For other Chinese characters on the cover, English translations are now provided in White font above each. While the late two projects ("2. Immunostimulatory activities of plant DNA"; "3. Characterization of binding sites recognized by anti-DNA antibodies") had been published in journals [12, 13], the first one ("Immunological memory mediated by antigen present cells") was not.

Zhou LJ, Tedder TF A distinct pattern of cytokine gene expression by human CD83+ blood dendritic cells. Blood 1995;86(9):3295-3301

题外话

树突状细胞介导免疫记忆，

这一设想的酝酿、成熟、付诸实施及获得初步结果，

与我女儿作为一个生命从无到有的发展几乎是完全同步的。

在同事们的调侃中，我采用的他们的建议，给我女儿取名为

“舒忆”（树忆）

希望女儿懂事以后，即使根本不懂树突状细胞为何物，也能理解我

的良苦用心，而不是嘲笑一个科技工作者的自作多情。

**Figure 2. Statement about naming the baby Shuyi after dendritic-cell mediated immunological memory.** This is a picture of the inner page in Wang's report telling the connection of Wang's first daughter with the hypothetical memory dendritic cells story. The Chinese wordings read as: "Digression. The hypothesis of dendritic cells mediating immunological memory was conceived, refined, tested, and gained preliminary supporting data in about the same pace with my daughter to be conceived, to develop and to be born to this family. Taking my colleagues' joking seriously, I decided to name my daughter Shuyi. I wish that when she grows up she will understand but not sneer at my being sentimental as a scientist, even if she might not know what dendritic cells are at all." Shuyi Wang is now a senior student in National Taiwan University of Technology majoring in Corporate Management. Luckily to me, she has been proud of her name and was excited about Hole's publication [7].